\documentclass[prl,twocolumn,amsfonts,nofootinbib,showpacs]{revtex4}
\usepackage{epsfig,bm,amsmath,color}
\usepackage[T1]{fontenc}

\newcommand{\be}{\begin{equation}}
\newcommand{\ee}{\end{equation}}
\newcommand{\bea}{\begin{eqnarray}}
\newcommand{\eea}{\end{eqnarray}}
\newcommand{\ba}{\begin{array}}
\newcommand{\ea}{\end{array}}
\newcommand{\al}{\alpha}
\newcommand{\dH}{\dot{H}}
\newcommand{\vx}{\mathbf{x}}
\newcommand{\vk}{\mathbf{k}}
\newcommand{\mH}{\mathcal{H}}
\newcommand{\Ji}{\mathcal{\chi}}

\newcommand{\Ro}{R_{(0)}}
\newcommand{\dRo}{\dot{R}_{(0)}}
\newcommand{\ddRo}{\ddot{R}_{(0)}}

\begin{document}
\preprint{}
\title{Linearized Treatment of Scalar perturbations in the Asymptotic
  Cosmological Model}
\author{J.L. Cort\'es}
\email{cortes@unizar.es}
\affiliation{Departamento de F\'{\i}sica Te\'orica,
Universidad de Zaragoza, Zaragoza 50009, Spain}
\author{J. Indur\'ain}
\email{indurain@unizar.es}
\affiliation{Departamento de F\'{\i}sica Te\'orica,
Universidad de Zaragoza, Zaragoza 50009, Spain}

\begin{abstract}
In this paper the implications of a recently proposed
phenomenological model of cosmology, the Asymptotic Cosmological
Model (ACM), on the behavior of scalar perturbations are
studied. Firstly we discuss new fits of
the ACM at the homogeneous level, including fits to the Type Ia
Supernovae UNION dataset, first CMB peak of WMAP5 and BAOs. The
linearized equations of scalar perturbations in the FRW metric are
derived. A simple model is used to compute the CMB
temperature perturbation spectrum. The results are compared with the
treatment of perturbations in other approaches to the problem of the
accelerated expansion of the universe.
\end{abstract}
\pacs{98.80.-k, 04.50.Kd, 95.36.+x}

\maketitle

\section{Introduction}
According to General Relativity (GR), if the universe is filled with the particles
of the Standard Model of particle physics, gravity should lead to a
decelerated expansion of the universe. However, in 1998 two
independent evidences of present accelerated expansion were presented
\cite{Riess1,Perlmutter1} and later confirmed by different
observations \cite{KOBE, WMAP,LSS}.

There is no compelling explanation for this cosmic acceleration,
but many intriguing ideas are being explored. These ideas can be classified
into three main groups: new exotic sources of the gravitational field with
large negative pressure \cite{DE} (Dark Energy), modifications of
gravity at large scales \cite{lemg} and rejection of the spatial
homogeneity as a good approximation in the description of the present
universe \cite{DEwoDE}.

Different models (none of them compelling) for the source responsible
of acceleration have been considered. Einstein
equations admit a cosmological constant $\Lambda$, which can be realized as
the stress-energy tensor of empty space. This $\Lambda$ together with Cold Dark
Matter, Standard Model particles and General Relativity form the current
 cosmological model, $\Lambda$CDM. However, quantum field theory predicts a
value of $\Lambda$ which is 120 orders of magnitude higher than observed.
Supersymmetry can lower this value 60 orders of magnitude, which is
still ridiculous \cite{Weinberg}.
In order to solve this paradox, dynamical Dark Energy models have been proposed.

This has also lead to explore the possibility that cosmic acceleration arises
from new gravitational physics. Again here several alternatives for a
modification of the Einstein-Hilbert action at large and small
curvatures \cite{Faraoni:2008mf}, or even higher
dimensional models \cite{Deffayet, DGP}, producing an accelerated
expansion have been identified.
All these analyses include an {\it ad hoc} restriction to actions involving simple
functions of the scalar curvature and or the Gauss-Bonnet tensor.
This discussion is sufficient to establish the point that cosmic acceleration can
be made compatible with a standard source for the gravitational field but it is
convenient to consider a more general framework in order to make a systematic
analysis of the cosmological effects of a modification of general relativity.

The Asymptotic Cosmological Model (ACM) was presented
\cite{Cortes:2008fy} as a strictly phenomenological
generalization of the Standard Cosmological Model including a
past and a future epoch of accelerated expansion. It follows from the
assumptions that GR is not a fundamental theory, but only a good
approximation when the Hubble rate $H$
  is between but far away from two
fundamental scales $H_-$ and $H_+$, which act as bounds on
$H$. A general covariant metric theory of gravity without
spatial curvature is assumed. The model is well defined in the
homogeneous approximation
and includes $\Lambda$CDM as a particular case.

In next section we review the ACM and we provide new fits
to the Type Ia Supernovae UNION dataset, first acoustic peak of CMB of
WMAP5 and BAOs. In the third section we
derive the linearized equations of the scalar perturbations of the
metric, following from
general covariance and a single new assumption on the
  perturbations. In the fourth section
we will consider how to solve the system of equations for adiabatic
perturbations in a given fluid. In the fifth section we derive the
CMB spectrum in the ACM.  In the sixth
  section we will compare the treatment of the scalar perturbations in
the ACM with other models
  which are equivalent in the homogeneous approximation. The last
section is devoted to the summary and conclusions.

\section{The Asymptotic Cosmological Model: Homogeneous Background}

The Asymptotic Cosmological Model (ACM) was introduced in
Ref. \cite{Cortes:2008fy}. In this model the universe is filled with
photons and neutrinos (massless particles), baryons (electrically
charged massive particles) and Dark Matter (electrically neutral
massive particles), but General Relativity (GR) is only a good
approximation to the gravitational interaction in a certain range of
the Hubble rate $H$, between but far from its
two bounds, $H_-$ and $H_+$. General Covariance and the absence of
spatial curvature are assumed.

The gravitational part of the action might include derivatives of the
metric of arbitrarily high order, and therefore arbitrarily high
derivatives of the scale factor should appear in the Friedman
Equations. We should start by considering a generalized first Friedman
equation
\be
8\pi G \rho(t) = 3 f(H(t),\dot{H}(t),...,H^{(n)}(t),...) \label{fh} \, ,
\ee
and the corresponding equation for the pressure will be derived using the continuity
equation
\be
\dot{\rho} = -3 H(\rho + p)\, . \label{cont}
\ee

However, as the resulting differential equations should be solved and only one of
its solutions deserves interest (the one describing the
evolution of the universe), we can use the one to one
correspondence between time $t$ and Hubble parameter $H$ (assuming $\dot{H} < 0$)
to write the modified First Friedman Equation as a bijective map linking
the total energy density $\rho$ with $H$ in our universe
\be
8\pi G \rho = 3 g(H) \label{gh} \, .
\ee
The use of the continuity equation \eqref{cont} enables us to write the modified
Friedman Equation for the pressure evaluated at the solution
corresponding to the cosmic evolution

\be
-8\pi G p = 3 g(H) + g'(H)\dot{H}/H \label{ph} \, .
\ee

The evolution of the universe will be determined by the concrete form of
the function $g(H)$, which we assume to be smooth. However, the most significant
features of the evolution at a given period can be described by some
simple approximation to $g(H)$ and the matter content. Those are a
pole at $H = H_+$ of order $\al_+$ and a zero of order $\al_-$ at $H =
H_-$. The energy density of the universe has a contribution from
both massless particles (radiation) and massive ones (matter).

For the sake of simplicity, the history of the universe can be then divided into
three periods. In the first period, $H \gg H_-$ so the effect of the lower bound
can be neglected and the universe is radiation dominated. The universe undergoes
(and exits from) an accelerated expansion which we call inflation. The simplest
parameterization is

\be
g(H) \approx H^2 \left( 1 - \frac{H^2}{H_+^2} \right)^{-\al_+} \, . \label{past}
\ee

In the second period we cannot neglect neither the effect of radiation nor the
effect of nonrelativistic matter, but $H_+ \gg H \gg H_-$. Then GR
offers a good description
of the gravitational interaction, $g(H) \approx H^2$ in this region, and the
universe performs a decelerated expansion.

In the third period, $H_+ \gg H$ so the effect of the upper bound can be
neglected, and the universe is matter dominated. This period corresponds to
the present time in which the universe also undergoes an accelerated expansion. The
simplest choice is

\be
g(H) \approx H^2 \left( 1 - \frac{H_-^2}{H^2} \right)^{\al_-} \, . \label{future}
\ee

We will assume that the transitions between these three periods are smooth
and that the details about these transitions are unimportant.

This model preserves the successes of the Cosmological Standard Model, while
giving a description of the early accelerated expansion
(inflation) and of the present one, including $\Lambda$CDM as a particular case
($\al_-= 1$).

Without any knowledge of the evolution of perturbations in this model, the
background evolution at late times can be used to fit the Type Ia Supernovae
UNION dataset \cite{Kowalski:2008ez}, the first acoustic peak in the Cosmic
Microwave (CMB) Background \cite{Dunkley:2008ie}, and the Baryon Acoustic
Oscillations (BAOs) \cite{Eisenstein:2005su}, via the parameters $H_0$
(the present Hubble parameter), $H_-$ and $\al_-$.

We use Monte Carlo Markov Chains to explore the likelihood of the fit of the
supernovae UNION dataset \cite{Kowalski:2008ez} to the ACM. The dataset provides
the luminosity distance
\be
d_l(z) = c(1+z)\int_0^z dz'/H(z')
\ee
and redshift of 307 supernovae. A $\chi^2$ analysis have been performed, where
$\chi^2$ has been marginalized over the nuisance parameter $H_0$ using the method
described in \cite{Wang}. The resulting parameter space is spanned by the values
of $\alpha_-$ and $H_-/H_0$ (FIG. 1).

\begin{figure}
 \centerline{\includegraphics[scale=0.6, width=8cm]{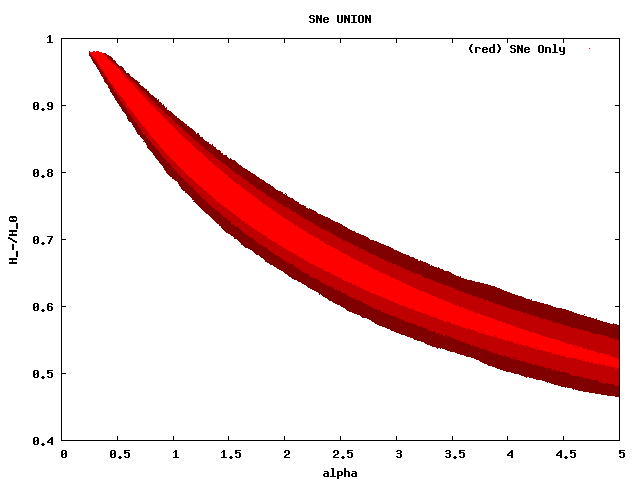}}
\label{U}
\caption{Confidence regions in parameter space of the Asymptotic
  Cosmological Model (ACM) from the fit of the supernovae UNION dataset
   without priors at $1\sigma$, $2\sigma$ and $3\sigma$. The $\Lambda$CDM is
  inside the 1$\sigma$ region. The best fit to ACM lies in
  $\alpha_- = 0.35$, $H_-/H_0 = 0.97$
($\chi^2 = 310.5$) in contrast to the best fit to $\Lambda$CDM, which lies in
$\alpha_- \equiv 1$ , $H_-/H_0 = 0.84$
($\chi^2 = 311.9$).}

\end{figure}

The constraints from the CMB data follow from the reduced distance to the surface
of last scattering at $z=1089$. The reduced distance $R$ is often written as
\be
R = \Omega_m^{1/2} H_0 \int_0^{1089} dz/H(z) \, .
\ee
The WMAP-5 year CMB data alone yield $R_0 = 1.715\pm 0.021$ for a fit assuming a
constant equation of state $\omega$ for the dark energy \cite{Komatsu:2008hk}. We
will take this value as a first approximation to the fit assuming ACM.  We can
define the corresponding $\chi^2$ as $\chi^2 =  [(R-R_0)/\sigma_{R_0}]^2$, and find
the confidence regions of the joint constraints (FIG. 2).

\begin{figure}[htbp]
 \centerline{\includegraphics[scale=0.6, width=8cm]{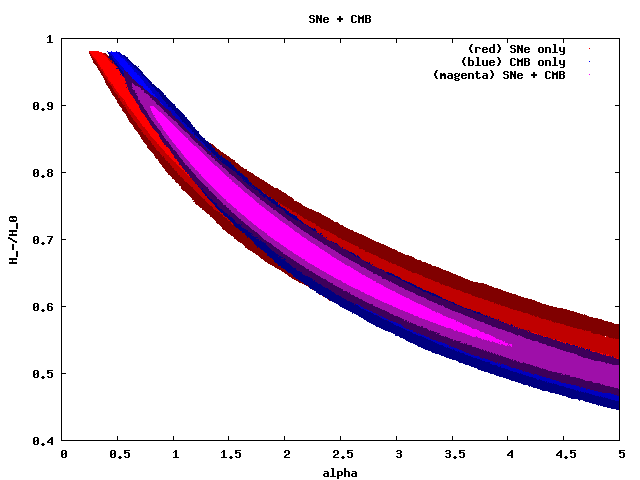}}
\label{UW}
\caption{Confidence regions in parameter space of the Asymptotic
  Cosmological Model (ACM) from the fit of the supernovae UNION
  dataset (red), of the distance to the surface of last scattering from WMAP-5
  (blue)  and from the joint fit (magenta) at $1\sigma$, $2\sigma$ and $3\sigma$.
  Significantly, the $\Lambda$CDM is still inside the 1$\sigma$ region, unlike in
  our previous study. This is due to the change in the Supernovae dataset. The
  best fit to ACM lies in $\alpha_- = 1.50$,
  $H_-/H_0 = 0.77$ ($\chi^2 = 312.5$) in contrast to the best fit to $\Lambda$CDM,
  which lies in $\alpha_- \equiv 1$, $H_-/H_0 = 0.86$
  ($\chi^2 = 313.2$).}
\end{figure}

BAO measurements from the SDSS data provide a
constraint on the distance parameter
$A(z)$ at redshift $z=0.35$,
\be
A(z)= \Omega_m^{1/2} H_0 H(z)^{-1/3} z^{-2/3} \left[\int_0^z dz'/H(z')\right]^{2/3} \, .
\ee
Ref.~\cite{Eisenstein:2005su} gives $A_0 = 0.469 \pm 0.17$. We can define the
corresponding $\chi^2$ as $\chi^2 =  [(A(z\,=\,0.35)-A_0)/\sigma_{A_0}]^2$ . The
confidence regions resulting from adding this constraint are shown in FIG. 3.

\begin{figure}[htbp]
 \centerline{\includegraphics[scale=0.6, width=8cm]{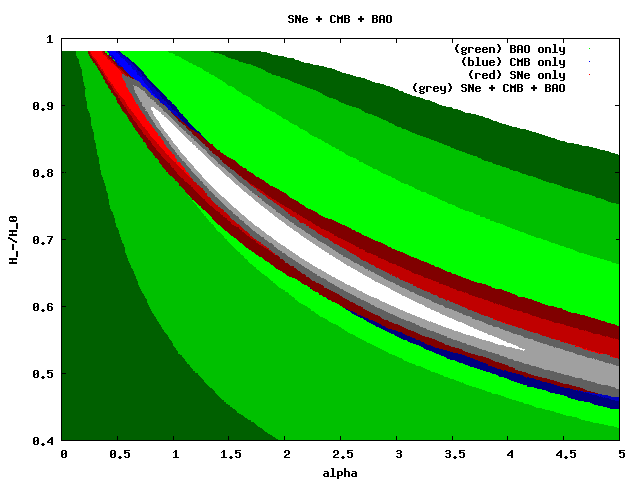}}
\label{UWB}
\caption{Confidence regions in parameter space of the Asymptotic
  Cosmological Model (ACM) from the fit of the supernovae UNION dataset (red),
  of the distance to the surface of last scattering from WMAP-5 (blue), of
  the Baryon Acoustic Oscillations peak (green) and from the joint fit
  (grey) at $1\sigma$, $2\sigma$ and $3\sigma$. Measurements of the
  BAOs peak do not add significant information with their present
  precision. The best fit to ACM lies in
  $\alpha_- = 1.50$, $H_-/H_0 = 0.77$
  ($\chi^2 = 312.7$) in contrast to the best fit to $\Lambda$CDM, which lies
  in $\alpha_- \equiv 1$, $H_-/H_0 = 0.86$
  ($\chi^2 = 313.7$).}
\end{figure}

In our previous work the use of the supernovae Gold dataset
\cite{Riess2} and of the WMAP-3 data \cite{WMAP3y} led us to the
conclusion that the $\Lambda$CDM was at $3\sigma$ level in the
parameter space of the ACM. The position of the confidence regions
seems to depend very tightly on the dataset that is being
used. However, the value of the combination of parameters
\be
\Omega_m \equiv \left(1-\frac{H_-^2}{H_0^2}\right)^{\al_-} \label{omegam}
\ee
does not depend much neither on the value of $\al_-$ or the
dataset used (FIG. 4).

Moreover, we can conclude from the figures that
BAO's do not provide much information in order to constrain the
confidence regions of the ACM, unlike in other models such as
$\Lambda$CDM with nonzero spatial curvature.

\begin{figure}[htbp]
 \centerline{\includegraphics[scale=0.6, width=8cm]{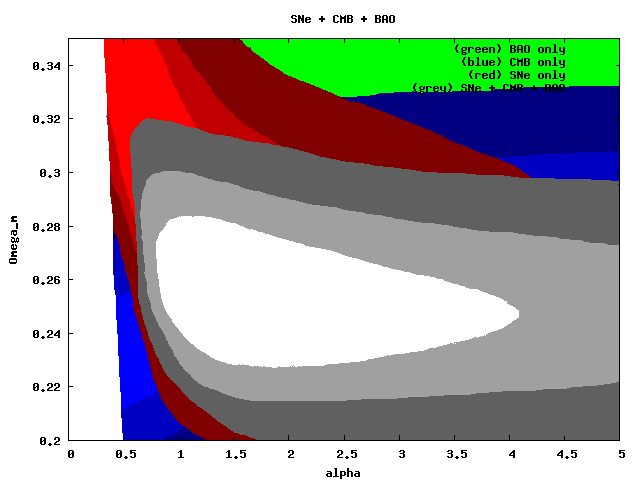}}
\label{omega}
\caption{Confidence regions in parameter space of the Asymptotic
  Cosmological Model (ACM) from the fit of the supernovae UNION dataset (red),
  of the distance to the surface of last scattering from WMAP-5
  (blue), of the Baryon Acoustic Oscillations peak (green) and from
  the joint fit (grey) at $1\sigma$, $2\sigma$ and $3\sigma$. The plot
  shows $\alpha_-$ in the horizontal axis and $\Omega_m$ in the
  vertical axis. Noticeably, $\Omega_m = 0.26 \pm 0.04$ almost
  independently of the value of $\alpha_-$.}
\end{figure}

\section{Scalar Perturbations}

The lack of an action defining the ACM is a serious obstacle in
the derivation of the equations governing the behavior of the
perturbations. Given a background behavior described by the ACM,
what can be said about the evolution of perturbations on top of this
background? We will find that general covariance together with an
additional assumption fixes completely the set of equations for the
scalar perturbations in the linearized approximation and in the region
close to $H_-$.

The key point will be the following. Knowing the exact Friedman
equations in the homogeneous approximation gives us a clue on the form
of the equations for the scalar perturbations. In particular, we can
formally describe perturbations over the FRW metric which do not depend
on the spatial coordinates, $\{\phi(\vx,t)=\phi(t),
\psi(\vx,t)=\psi(t)\}$. Thus the perturbed
metric becomes the FRW metric
written in a new coordinate frame. With a change
of coordinates one can derive, starting from the Friedman equations,
the terms containing only time derivatives of the scalar
perturbations. Next an assumption on the validity of the GR
description of scalar perturbations when $H_-\ll H \ll H_+$,
together with the relations, valid for any general covariant theory,
between terms with time derivatives and those involving spatial
derivatives, allow to derive the evolution of scalar perturbations.

After this short sketch, we will perform the derivation of the
equations for the scalar perturbations in detail. We can write the
metric of spacetime with scalar perturbations in the Newtonian gauge,
\be
ds^2 = (1+2\phi(\vx,t))dt^2 - a^2(t) (1-2\psi(\vx,t)) d\vx^2 \, ,\label{newt}
\ee
and the stress-energy tensor of the source fields will be
\be
\ba{ccl}
T^0_0 & = & \rho_{(0)}(t) + \delta \rho(\vx,t) \, , \\
T^0_i & = & (\rho_{(0)}(t) + p_{(0)}(t)) \partial_i \theta(\vx,t) \, ,\\
T^i_j & = & -(p_{(0)}(t) +\delta p(\vx,t))\delta^i_j\,+\,\partial_i\partial_j\Pi(\vx,t)\, ,
\ea \label{set}
\ee
where $\theta$ is the velocity potential of the fluid, $\Pi$ is the
anisotropic stress tensor (shear) potential, $\delta\rho$ and $\delta
p$ are small perturbations on top of the background homogeneous
density $\rho_{(0)}$ and pressure $p_{(0)}$, respectively, and
$\phi\sim\psi\ll 1$. We have used the Newtonian gauge because the
gauge invariant scalar perturbations of the metric ($\Phi$ and $\Psi$)
and gauge invariant perturbations in the stress energy tensor
coincide with the perturbations explicitly written in this gauge.

The general form of the equations for the perturbations in a metric
theory with arbitrarily high derivatives is

\begin{widetext}
\begin{eqnarray}
8 \pi G \delta\rho & = & \sum_{n=0}^\infty \sum_{m=0}^\infty
\left[\frac{a_{nm}}{H^{2n+m-2}}\left(\frac\Delta{a^2}\right)^n
  \partial_t^m \phi +
  \frac{b_{nm}}{H^{2n+m-2}}\left(\frac\Delta{a^2}\right)^n
  \partial_t^m \psi \right] \label{rho1} \, ,\\
8 \pi G (\rho_{(0)} + p_{(0)}) \partial_i \theta & = &
\partial_i\sum_{n=0}^\infty \sum_{m=0}^\infty
\left[\frac{c_{nm}}{H^{2n+m-1}}\left(\frac\Delta{a^2}\right)^n
  \partial_t^m \phi +
  \frac{d_{nm}}{H^{2n+m-1}}\left(\frac\Delta{a^2}\right)^n
  \partial_t^m \psi \right] \label{vel1} \, ,\\
8 \pi G (-\delta p \,\delta^i_j + \partial_i \partial_j \Pi) & = &
\delta^i_j\sum_{n=0}^\infty \sum_{m=0}^\infty
\left[\frac{e_{nm}}{H^{2n+m-2}}\left(\frac\Delta{a^2}\right)^n
  \partial_t^m \phi +
  \frac{f_{nm}}{H^{2n+m-2}}\left(\frac\Delta{a^2}\right)^n
  \partial_t^m \psi \right] \nonumber \\
& & +\sum_{n=0}^\infty \sum_{m=0}^\infty \frac 1{a^2}(\delta^i_j
\Delta -\partial_i
\partial_j)\left[\frac{g_{nm}}{H^{2n+m}}\left(\frac\Delta{a^2}\right)^n
  \partial_t^m \phi +
  \frac{h_{nm}}{H^{2n+m}}\left(\frac\Delta{a^2}\right)^n \partial_t^m
  \psi \right] \label{p1}  \, .
\end{eqnarray}
\end{widetext}

The coefficients $a_{nm},...,h_{nm}$ are adimensional functions of the
Hubble parameter an its time derivatives, and can be turned into
functions of just the Hubble parameter using the bijection explained
in the beginning of the previous section. We have mentioned we are
going to be able to determine exactly the terms with only time derivatives, that
is, the precise form of the coefficients $a_{0m},b_{0m},e_{0m}$ and
$f_{0m}$. This shows that there is much freedom of choosing a
covariant linearized theory of cosmological perturbations, even for a
given solution of the homogeneous equations.

However, under a single assumption, it is possible to greatly reduce
this freedom. The assumption is that the standard linearized equations
of the perturbations of General Relativity are effectively recovered
in the limit $H_+ \gg H \gg H_-$ for all the Fourier modes with $H_+
\gg k > H_-$ (subhorizon modes in the present time).

Let us work out the implications of this assumption. In the equation
for the perturbation of the energy density in General relativity,
\be
8\pi G \delta \rho  =  -6 H^2 \phi -6 H \dot \psi +\frac{2}{a^2}\Delta \psi\, ,
\ee
there is only a term proportional to $\Delta \psi$. However, for
subhorizon modes ($k\gg H$) the dominant terms are those with the
highest number of spatial derivatives. If we demand
the terms with spatial derivatives of order greater than $2$ not to
spoil the behavior of these modes in the period in which GR is a good
approximation, they must be negligible at least for the observable modes.

For instance, they should
be negligible for the modes responsible of the acoustic peaks of the
CMB spectrum, and therefore for the modes which have entered the
horizon after recombination (which have even
lower $k$). It is possible that these terms
are suppressed by inverse powers of the UV scale
$H_+$, becoming irrelevant for
current tests of gravity, although they may be relevant for the
physics of quantum fluctuations in the very early universe.

Therefore, at times when $H\ll H_+$, the equations can be approximated by

\begin{widetext}
\begin{eqnarray}
8 \pi G \delta\rho & = & \sum_{m=0}^\infty
\left[\frac{a_{0m}}{H^{m-2}}\partial_t^m \phi +
  \frac{b_{0m}}{H^{m-2}}\partial_t^m \psi +
  \frac{a_{1m}}{H^{m}}\frac\Delta{a^2} \partial_t^m \phi +
  \frac{b_{1m}}{H^{m}}\frac\Delta{a^2}\partial_t^m \psi \right]
\label{rho2} \, ,\\
8 \pi G (\rho_{(0)} + p_{(0)}) \partial_i \theta & = &
\partial_i\sum_{m=0}^\infty \left[\frac{c_{0m}}{H^{m-1}}\partial_t^m
  \phi + \frac{d_{0m}}{H^{m-1}}\partial_t^m \psi \right] \label{vel2}
\, ,\\
8 \pi G (-\delta p \,\delta^i_j + \partial_i \partial_j \Pi) & = &
\delta^i_j \sum_{m=0}^\infty \left[\frac{e_{0m}}{H^{m-2}}\partial_t^m
  \phi + \frac{f_{0m}}{H^{m-2}}\partial_t^m \psi \right] \nonumber \\
& & +\sum_{m=0}^\infty \frac 1{a^2}(\delta^i_j \Delta -\partial_i
\partial_j)\left[\frac{g_{0m}}{H^{m}} \partial_t^m \phi +
  \frac{h_{0m}}{H^{m}}\partial_t^m \psi \right] \label{p2} \, ,
\end{eqnarray}
\end{widetext}
where general covariance has been used to exclude the terms with
coefficients $e_{1m}$, $f_{1m}$ which are not compatible with
(\ref{rho2}), (\ref{vel2}).

We must also take into account that the modes which are responsible of
the acoustic peaks of the CMB undergo a phase in which they oscillate
as sound waves, i.e.: $\partial_t\phi~\sim~k\phi$. In order to
explain the acoustic peaks of the spectrum of the CMB it is required
that the acoustic oscillations of the modes of the gravitational
potentials, $\phi_\vk$ and $\psi_\vk$, which lead to the acoustic
peaks of the CMB spectrum, have a frequency $\sim k$. When radiation
dominates and we consider modes well inside the horizon, two of the
solutions of the system of differential equations of arbitrary order
have this property \footnotemark[1].

However, in order for nondecaying superhorizon modes to evolve
into oscillating subhorizon modes with frequency $\sim k$, a fine-tuning of
coefficients is required unless the equations are of order two in time
derivatives. Then the acoustic peaks in the CMB spectrum are
reproduced when terms with more than two derivatives in the equations
for the  perturbations can be neglected. Then
\begin{widetext}
\begin{eqnarray}
8 \pi G \delta\rho & = & \sum_{m=0}^2
\left[a_{0m}H^{2-m}\partial_t^m \phi +
  b_{0m}H^{2-m}\partial_t^m \psi \right]+
a_{10}\frac\Delta{a^2}\phi + b_{10}\frac\Delta{a^2}\psi  \label{rho3}
\, ,\\
8 \pi G (\rho_{(0)} + p_{(0)}) \partial_i \theta & = &
\partial_i\sum_{m=0}^1 \left[c_{0m}H^{1-m}\partial_t^m \phi +
  d_{0m}H^{1-m}\partial_t^m \psi \right] \label{vel3} \, ,\\
8 \pi G (-\delta p \,\delta^i_j + \partial_i \partial_j \Pi) & = &
\delta^i_j \sum_{m=0}^2 \left[e_{0m}H^{2-m}\partial_t^m \phi
  + f_{0m}H^{2-m}\partial_t^m \psi \right] \nonumber \\
& & +\frac 1{a^2}(\delta^i_j \Delta -\partial_i
\partial_j)\left[g_{00}\phi + h_{00}\psi \right] \label{p3} \, ,
\end{eqnarray}
\end{widetext}

\footnotetext[1]{
As we will see
below, if the term with the highest time derivative is of order
$D$,  $a_{1,D-2} = -e_{0D}$ as a consequence of general
covariance. Well inside the horizon these are the dominant terms
appearing in the wave equation resulting when calculating the
adiabatic perturbations in the radiation domination
epoch.}

\noindent and the arbitrariness in the evolution equations for the scalar
perturbations has been reduced to twenty undetermined dimensionless
coefficients at this level.

The equations \eqref{rho3},\eqref{vel3},\eqref{p3} are in principle
valid for any perturbation mode $\{\phi_\vk,\psi_\vk\}$ as long as $k\ll H_+$.
In particular, it must be valid for the mode $\vk = \mathbf{0}$, which
corresponds formally to a perturbation with no spatial dependence. In practice,
we will be able to neglect the spatial dependence of perturbation whose spatial
dependence is sufficiently smooth, i.e.: its wavenumber $k$ is sufficiently low.
In GR it suffices for a mode to be superhorizon $k\ll H$ in order to neglect
its spatial dependence in a first approximation.

If we consider the FRW metric perturbed by one of these modes we have,
\be
ds^2 = (1+2\phi(t))dt^2 - a^2(t) (1-2\psi(t)) d\vx^2 \, . \label{superh}
\ee

>From now on the dot will represent derivative with respect to the cosmic time.
By means of the invariance under time reparameterizations we can introduce
a new time variable $dt'= (1+\phi(t))dt$, a new scale factor
$a'(t')=a(t)(1-\psi(t))$ and a new energy density
$\rho'(t')=\rho_0(t)+\delta\rho(t)$
leading us back to the ACM in a homogeneous background. The Hubble
rate $H = \frac{\dot a}{a}$ and its time derivatives $H^{(n)}$ will change as
\be
\begin{array}{l}
H^{(n)'} =
\frac{d^n}{dt^{'n}}\left[\frac{1}{a'(t')}\frac{da'}{dt'}\right] =
  H^{(n)} +\delta H^{(n)}, \\
\delta H^{(n)} = -\psi^{(n+1)}
-\sum_{m=0}^n
\left(\ba{c}
n+1 \\ m+1
\ea
\right) H^{(n-m)}\phi^{(m)}.\label{Hn}
\end{array}
\ee
Noticeably, this procedure must be applied to the generalized Friedman
equation before one of its homogeneous solutions is used to write the
time variable $t$ as a function of $H$. Therefore we should start by
considering a generalized first Friedman equation in primed
coordinates
\be
8\pi G \rho'(t') = 3 f(H'(t'),\dot{H}'(t'),...,H^{(n)'}(t'),...) \label{fh'} \, ,
\ee
and the corresponding equation for the pressure will be derived using
the continuity equation. The linearized equations for the sufficiently smooth
 scalar perturbations in the Newtonian gauge~(\ref{superh}) can then be
derived with the help of~(\ref{Hn}). The result is

\begin{eqnarray}
8\pi G \delta \rho(t) & = & \sum_{n = 0}^\infty g_{\vert n}(H(t))\delta H^{(n)}(t),
\label{rho4} \\
-8\pi G \delta p & = & \frac 1H \sum_{n = 0}^\infty\sum_{m = 0}^\infty
g_{\vert nm}(H)H^{n+1}\delta H^{(m)}\nonumber\\
& &+\sum_{n = 0}^\infty g_{\vert n}(H) \left(\frac{\delta
  H^{(n+1)}}H-\frac{H^{(n+1)}}{H^2}\delta H \right)\nonumber\\
& &+ 3\sum_{n = 0}^\infty g_{\vert n} (H)\delta H^{(n)},\label{p4}
\end{eqnarray}
where
\begin{eqnarray}
 g_{\vert n} (H) & = & \left[ \frac{\partial f}{\partial
     H^{(n)}}\right]_{H^{(n)} = H^{(n)}(t(H))}\, ,\\
g_{\vert nm}(H) & = & \left[ \frac{\partial^2 f}{\partial
     H^{(n)}\partial H^{(m)}}\right]_{H^{(n)} = H^{(n)}(t(H))} \, ,
\end{eqnarray}
and the time dependence of $H$ is derived from the
homogeneous evolution \eqref{gh}. This procedure fixes exactly the
coefficients $a_{0m},b_{0m},e_{0m}$ and $f_{0m}$ in
\eqref{rho2},\eqref{vel2},\eqref{p2}, i.e. all the terms with no
spatial derivatives in the equations for the scalar perturbations, as
functions of $f(H,\dot{H},...)$ and its partial derivatives.

We can now count the number of functional degrees of freedom in the
superhorizon modes of the linearized theory. Let us assume
that there is a maximum number $D$ of time derivatives in the equations
for the perturbations~\footnotemark[2].\footnotetext[2]{We are
  forgetting here the restriction on the
number of derivatives required in order to reproduce the acoustic peaks of
CMB.} The number of derivatives, with the use of
\eqref{Hn},\eqref{rho4},\eqref{p4}, fixes $f$ to be a function of at
most $H^{(D-2)}$. When particularized to a solution of the homogeneous
Friedman equations, $f$ and its first and second partial derivatives
become functions of the Hubble rate $H$: $g(H)$, $g_{\vert n}(H)$ and
$g_{\vert nm}(H)$ respectively. That makes $1+(D-1)+D(D-1)/2=D(D+1)/2$
functional degrees of freedom. The first derivative of $g(H)$ can be
written in terms of the $g_{\vert n}(H)$, and the first derivative of the
later can be also written in terms of the $g_{\vert nm}(H)$. That makes $1 +
(D-1)$ conditions, so the result depends on $D(D-1)/2$ independent
functions of the Hubble parameter $H$ in the terms without any spatial
derivatives.

Notice however that if we restrict the number of derivatives appearing
in the equations for the perturbations to two as in
Eqs.~\eqref{rho3},\eqref{vel3},\eqref{p3}, then
Eqs.~\eqref{Hn},\eqref{rho4},\eqref{p4} tell us that we must consider
just functions $f(H,\dot H, \ddot H,...)\, =\, f(H)\, = \, g(H)$, at
least as an approximation at times $H\ll H_+$ and modes $k \ll
H_+$. Therefore the number of functional degrees of freedom in the
terms with no spatial derivatives is just one: the homogeneous
evolution $f(H) = g(H)$.

Our assumption might be relaxed and we could impose that general
relativity should be valid for all the modes $H_-<k<k_{obs}$ that have
been observed in the spectrum of CMB and matter perturbations. This
could lead to new terms in the equations for the perturbations
(suppressed not necessarily by the UV scale $H_+$) which have been
negligible for the observed modes but that could lead to ultraviolet
deviations from the spectrum derived in the general relativistic
cosmology which have not yet been observed. However, terms with more
than two spatial derivatives will be very tightly constrained by solar
system experiments.

In this article we will restrict ourselves to the system of second order
 differential equations (\ref{rho3}), (\ref{vel3}),
(\ref{p3}).
Let us now derive the linearized equations for
the rest of the modes under this assumption. In the $(t',\vx)$
coordinate system, the metric is Friedman Robertson Walker, and we
know that for this metric, we can use equation (\ref{gh}). Thus,
\be
8 \pi G \rho' = 3 g(H') \, ,
\ee
with
\be
H' = \frac 1 {a'} \frac{d a'}{d t'} = H(1-\phi)-\dot{\psi}
\ee
at linear order in perturbations. Therefore, we can deduce from
(\ref{gh}) that in the $(t,\vx)$ coordinate system
\begin{widetext}
\begin{eqnarray}
8 \pi G \delta\rho & = & -3 g'(H)\left(H \phi +\dot{\psi}\right)+
a_{10}\frac\Delta{a^2}\phi + b_{10}\frac\Delta{a^2}\psi  \label{rho5}
\, ,\\
8 \pi G (\rho_{(0)} + p_{(0)}) \partial_i \theta & = &
\partial_i\sum_{m=0}^1 \left[\frac{c_{0m}}{H^{m-1}}\partial_t^m \phi +
  \frac{d_{0m}}{H^{m-1}}\partial_t^m \psi \right] \label{vel5} \, ,\\
8 \pi G (-\delta p \,\delta^i_j + \partial_i \partial_j \Pi) & = &
\delta^i_j  \left[g'(H)\left(3 H \phi +3
  \dot{\psi}-\frac{\dot{H}}{H^2}\dot{\psi}+\frac{\dot{H}}H \phi
  +\dot{\phi} + \frac 1 H \ddot{\psi}\right) \right. \nonumber\\
& & \left. +\frac{\dot{H}}H g''(H)(H\phi+\dot{\psi}) \right] \nonumber \\
& & +\frac 1{a^2}(\delta^i_j \Delta -\partial_i
\partial_j)\left[g_{00}\phi + h_{00}\psi \right] \label{p5} \, ,
\end{eqnarray}
\end{widetext}
and the twelve coefficients of terms with no spatial derivatives
are fixed by the function $g(H)$ which defines the homogeneous cosmological
model.

The last requirement comes again from general covariance. If the
tensor $T_{\mu\nu}$ comes from the variation of a certain matter
action $S_m$ with respect to the inverse of the metric $g^{\mu\nu}$,
and $S_m$ is a scalar under general coordinate transformations, then
$T^\mu_\nu$ must be a 1-covariant, 1-contravariant divergenceless
tensor, i.e.  $\nabla_\mu T^\mu_\nu = 0$. As the stress-energy tensor
is proportional to $G^\mu_\nu$, the later must also be
divergenceless. In the linearized approximation $G^\mu_\nu =
G^\mu_{(0)\nu}+\delta G^\mu_\nu$ and the Christoffel symbols
$\Gamma^\lambda_{\mu\nu} = \Gamma^\lambda_{(0)\mu\nu}+\delta
\Gamma^\lambda_{\mu\nu}$  , and we can write the linearized version of
this requirement as

\be
\ba{c}
\delta G^0_{\mu\vert 0} +\delta G^i_{\mu\vert i} + \delta
G^\lambda_\mu \Gamma^\nu_{(0)\lambda\nu} - \delta G^\lambda_\nu
\Gamma^\nu_{(0)\lambda\mu} \\
+ G^\lambda_{(0)\mu} \delta\Gamma^\nu_{\lambda\nu}
- G^\lambda_{(0)\nu} \delta\Gamma^\nu_{\lambda\mu} = 0 \, .
\ea
\ee

Let us study this condition in order to see if we can further limit
the number of independent coefficients of
the equations for scalar perturbations. For $\mu = i$ the term
$\delta G^0_{i\vert 0}$ will give at most a term proportional to
$d_{01}\ddot{\psi}_{\vert i})$, which will not be present in other
terms except $\delta G^j_{i \vert j}$. This term will give at most a
term $g'\ddot{\psi}_{\vert i}/H$ which fixes the
exact value of $d_{01}$ for all $H \ll H_+$ . For the same reason,
the term $c_{01}\ddot{\phi}_{\vert i}$ in $\delta G^0_{i\vert 0}$
can not be canceled and the term $c_{00}H\dot{\phi}_{\vert i}$ can
only be canceled by the term $g'\dot{\phi}_{\vert i}$ in
$\delta G^j_{i \vert j}$, which means that $c_{01}=0$ and fixes
 the exact value of $c_{00}$ for all $H\ll H_+$. The value
of $d_{00}$ can be fixed in
terms of $d_{01}$ and the terms proportional to $\dot{\psi}$ in
(\ref{p5}). This fixes (\ref{vel5}) completely in terms of the
function describing the homogeneous evolution, $g(H)$:

\be
8 \pi G (\rho_0+p_0)\partial_i\bar{\theta}  =
\frac{g'(H)}{H}\partial_i \left (H \Phi + \dot{\Psi} \right )
\label{vdef} \, .
\ee

In the previous equation and from now on we will refer directly to the
gauge invariant counterparts of the variables in the Newtonian gauge,
which we will denote by $\Phi,\Psi,\bar{\theta},\bar{\delta\rho}$ and
$\bar{\delta p}$ ($\Pi$ is already gauge invariant). Now we can use
the divergenceless condition for $\mu=0$. The term $\delta G^0_{0\vert
  0}$ will give at most a term proportional to $a_{10}\Delta
\dot{\phi}$ and a term proportional to $b_{10}\Delta \dot{\psi}$,
which will not be present in other terms except $\delta G^j_{0 \vert
  j}$. This term will give at most a term proportional to $\Delta
\phi$ and a term proportional to $\Delta \dot{\psi}$. Therefore
$a_{10}=0$  and the value of $b_{10}$ is set completely in terms of
$g(H)$:

\be
8 \pi G \bar{\delta \rho} = -3 g'(H)(H \Phi +\dot{\Psi})+ \frac
{g'(H)}{a^2 H}\Delta \Psi
\label{rhodef} \, .
\ee

The complete knowledge of (\ref{rhodef}) and (\ref{vdef}) when $H \ll
H_+$ gives a complete knowledge of (\ref{p5}) in terms of $g(H)$ when
$H \ll H_+$:
\begin{widetext}
\be
\ba{ccl}
8 \pi G (\bar{\delta p} \, \delta_{ij} + \partial_i \partial_j \bar{\Pi})
& = &
\delta_{ij}\left\{
g'(H)\left[3 H \Phi +3
  \dot{\Psi}-\frac{\dot{H}}{H^2}\dot{\Psi}+\frac{\dot{H}}H \Phi
  +\dot{\Phi} + \frac 1 H \ddot \Psi \right]
+\frac{\dot H}H g''(H)\left[H \Phi + \dot \Psi \right]
\right\}\\
& & +\frac 1{a^2}\left(\delta_{ij}\Delta -\partial_i\partial_j\right)
\left( \frac{g'(H)}{2 H}(\Phi-\Psi) + \frac{\dot H}{2 H^3}(g'(H) - H
g''(H))\Psi \right)
\ea\label{pdef} \, ,
\ee
\end{widetext}
and we have finally a set of equations for the scalar perturbations
(\ref{rhodef}), (\ref{vdef}), (\ref{pdef}) which are determined by the
function $g(H)$ which defined the ACM in the homogeneous approximation.

Had we relaxed our assumption, we could work with the system of
equations~\eqref{rho1},\eqref{vel1},\eqref{p1} restricted to a maximum
number of derivatives $D$ as an approximation. Together with the
condition $G^\mu_{\nu;\mu} =0$, the system of equations for the
perturbations define a set of coupled differential equations for the
coefficients of the terms with at least one spatial derivative. The
set of equations coming from $G^\mu_{i;\mu}=0$ define the terms
proportional to $e_{mn}$ and $f_{mn}$ in \eqref{p1} as a function of the
terms in \eqref{vel1}. The set of equations coming from
$G^\mu_{0;\mu}$ define the rest of the terms in \eqref{p1} as a
function of the terms in \eqref{vel1} and \eqref{rho1}. Therefore, the
only freedom, for what concerns linearized perturbations, is that of
choosing the set of functions $\{a_{nm}, b_{nm}, c_{nm}, d_{nm}\}$,
with some of them fixed by the homogeneous dynamics \eqref{fh}.  An
interesting case is found if $D=4$ is imposed. This includes a
description of $f(R)$-theories \cite{Faraoni:2008mf}, bigravity
theories \cite{Damour:2002wu}, and other of the most studied modified
gravity theories. We will study further this case
in the subsection devoted to the comparison of the model with $f(R)$
theories.

The rhs of the equations derived are the equivalent to the components
of the linearized Einstein tensor derived in a general covariant
theory whose Friedman equation is exactly (\ref{gh}). Aside from
General Relativity with or without a cosmological constant, i.e. for
$\alpha_-\neq 1, 0$, it will be necessary to build an action depending
on arbitrarily high derivatives of the metric in order to derive a
theory such that both the equations in the homogeneous approximation
and the linearized equations for the scalar perturbations are of
finite order.

Noticeably, the nonlinear equations for the perturbations will include
derivatives of arbitrary order of the metric perturbations, with the
exceptional case of General Relativity with a cosmological
constant. This makes problematic the consistency of the ACM beyond the
linear approximation. This issue will be further studied in a future
work.

To summarize, what we have shown in this section is
that the behavior of scalar perturbations in a general covariant
theory is intimately connected to the background evolution.
Except in the very early universe, the
linearized equations for the scalar perturbations are determined by
the equations in the homogeneous approximation if one assumes that
there are no terms with more than two spatial derivatives. This
assumption could be relaxed in order to include more general theories.

There will be a subset of general covariant theories with a background
evolution given by (\ref{gh}) that verify these
equations for the perturbations. We may be able to distinguish among
these at the linearized level by means of the vector and tensor
perturbations.

\section{Hydrodynamical Perturbations}

Vector perturbations represent rotational flows which decay very
quickly in the General Relativistic theory. As we expect a small
modification of the behavior of perturbations just in the vicinity of
the lower bound $H_-$, we will assume that the vector perturbations
have decayed to negligible values when the scale $H_-$ begins to play
a role and therefore they will be ignored.

Tensor perturbations correspond to gravitational waves, which at the
present have been not observed, their effect being far beyond the
resolution of current observations.

Present measurements restrict their attention to perturbations in the
photon sector (CMB) and the matter sector (matter power spectrum, both
dark and baryonic). These observations can be computed taking into
account only the effect of scalar perturbations. Therefore, we can use
the result of the previous section to study the deviations from
$\Lambda$CDM in the spectrum of CMB. A comprehensive study of the
matter power spectrum predicted by the ACM would require a detailed
knowledge of the nonlinear regime, which we lack at present.

Let us start by deriving the behavior of scalar perturbations in a
universe whose gravitation is described by the ACM and filled with a
perfect fluid ($\Pi = 0$) in (\ref{set}). Therefore, (\ref{pdef}) with
$i\neq j$ fixes $\Phi$ as a function of $\Psi$,
\be
\Phi = \left\{1\,+\,\frac{\dot{H}}{2 H^2}\left(1 - \frac{H
  g''(H)}{g'(H)}\right)\right\}\Psi \label{Phi}\, .
\ee
Then, this expression can be used to substitute $\Phi$ in the
remaining three differential equations for $\bar{\delta\rho}$,
$\bar{\delta p}$ and $\bar{\theta}$. Given the equation of state of
the fluid $p = p(\rho, S)$, with $S$ the entropy density, the
perturbation of the pressure can be written as
\be
\bar{\delta p} = c_s^2 \bar{\delta \rho} + \tau \delta S \, \label{ppert},
\ee
where $c_s^2 \equiv (\partial p/\partial \rho)_S$ is the square of the
speed of sound in the fluid and $\tau \equiv (\partial p/\partial
S)_\rho$. Substituting the pressure and the energy density
perturbations for the corresponding functions of $\Psi$ and its
derivatives into (\ref{ppert}), we arrive at the equation for the
entropy density perturbation $\delta S$. The perturbations we are
interested in are adiabatic, i.e.: $\delta S = 0$ and therefore, the
equation for the entropy perturbations turns into an equation of
motion of the adiabatic perturbations of the metric. If we want to
consider the effect of the lower bound of the Hubble rate, $H_-$, on
the evolution of perturbations, we must consider a matter dominated
universe ($c_s^2 = 0$),
\be
\ba{c}
g'(H)\left[3 H \Phi +3
   \dot{\Psi}-\frac{\dot{H}}{H^2}\dot{\Psi}+\frac{\dot{H}}H \Phi
   +\dot{\Phi} + \frac 1 H \ddot{\Psi}\right] \\
+\frac{\dot{H}}H g''(H)\left[H\Phi+\dot{\Psi}\right] \; = \; 0 \, .
\ea \label{Psi}
\ee
This equation is a second order linear differential equation with
non-constant coefficients. Thus it is useful to work with the Fourier
components of the metric perturbation
$\Psi_k$. The equation of motion for the Fourier components is just a
second order linear ODE with non-constant coefficients, which will be
analytically solvable just for some simple choices of $g(H)$.
However, in general, it will be mandatory to perform a numerical
analysis.

\section{The CMB Spectrum in ACM versus $\Lambda$CDM}

The main observation that can be confronted with the predictions of
the theory of cosmological perturbations at the linearized level is
the spectrum of temperature fluctuations of the Cosmic Microwave
Background (CMB), from which WMAP has recorded accurate measurements
for five years \cite{Komatsu:2008hk}.

The temperature fluctuations $\frac{\delta T}T$ are connected to the
metric perturbations via the Sachs-Wolfe effect \cite{Sachs:1967er},
which states that, along the geodesic of a light ray
$\frac{dx^i}{dt}=l^i(1+\Phi+\Psi)$, characterized by the unit
three-vector $l^i$, the temperature fluctuations evolve according to

\be
(\frac \partial{\partial t} + \frac {l^i} a \frac \partial{\partial
  x^i})(\frac{\delta T}T+\Phi) = \frac \partial {\partial  t}(\Psi +
\Phi) \, .
\ee

Neglecting a local monopole and dipole contribution, taking
recombination to be instantaneous at a certain time and assuming
that the reionization optical depth is negligible, the present
temperature fluctuation of a distribution of photons coming from a
given direction of the sky $l^i$ can be related to the temperature
perturbation and metric perturbation $\Psi$ in the last scattering
surface plus a line integral along the geodesic of
the photons of the derivatives of $\Phi$ and $\Psi$ (Integrated Sachs
Wolfe effect, ISW).

These relations are purely kinematical and remain unchanged in the
ACM. In order to take into account all the effects involved in the
calculation of the CMB spectrum the code of CAMB \cite{CAMB} has been
adapted to the ACM. (see the appendix for details).

We have compared the late time evolution of the perturbations
described by the ACM \eqref{future} for several values of $\al_-$ (taking into
account that $\al_- = 1$ corresponds to $\Lambda$CDM) for constant
$H_0$ and $\Omega_m$ defined in Eq.~\eqref{omegam}. We have
assumed a Harrison-Zel'dovich scale invariant spectrum of scalar
perturbations as initial condition ($n_s = 0$) in the region in
which the universe is radiation dominated but $H\ll H_+$ . Thus the study of the
effect of the scale $H_+$ is postponed. We have taken $H_0 = 72 \pm 8\,
km/sMpc$ from the results of the Hubble Key Project
\cite{Freedman:2000cf} and $\Omega_m = 0.26 \pm 4$ from our previous
analysis of the evolution at the homogeneous level. The effect of ACM
as compared to $\Lambda$CDM is twofold (FIG. 5).

\begin{figure}[htbp]
 \centerline{\includegraphics[scale=0.6, width=9cm]{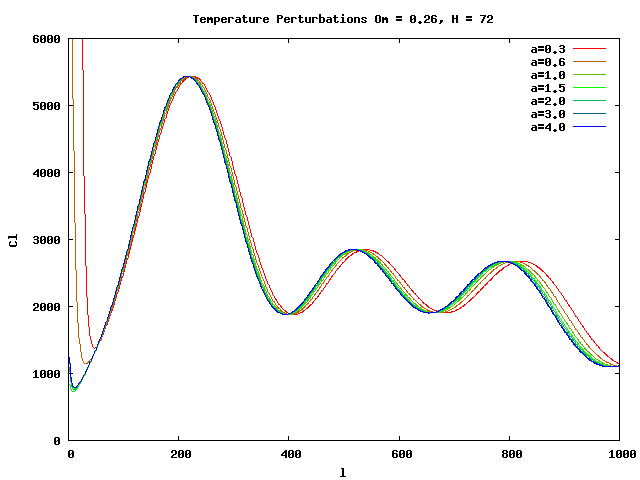}}
\label{spectrum}
\caption{Spectrum of temperature fluctuations of the Cosmic Microwave
  Background for $\Omega_m = 0.26$, $H_0 = 72 km/sMpc$ and varying
  $\alpha_- = 0.3$ (red), $0.6$, $1.0$, $1.5$, $2.0$, $3.0$, $4.0$
  (blue) . The lowest values of $\alpha_-$ show extreme ISW effects
  and therefore will be ruled out also by the CMB
  spectrum. Unfortunately, the cosmic variance masks the effect of
  $1.0 < \alpha_- < 4.0$ at large angular scales and therefore we can
  obtain little information about the ACM from the late ISW effect.}
\end{figure}

There is a shift in the peak positions due to the difference in the
distance to the last scattering surface for different values of
$\alpha_-$. This shift increases with the peak number. In particular,
more precise measurements of the position of the third peak could be
used to estimate the value of $\alpha_-$.

There is also an increase of the lower multipoles due to the late
Integrated Sachs-Wolfe effect, which is particularly extreme for
$\alpha_- \lesssim 1.0$ (which corresponds to $H_-\simeq H_0$ for
constant $\Omega_m$). This large deviation is clearly not present in
the experimental data. The much smaller deviation for $\alpha_-
\gtrsim 1.0$ is within the error bands due to the cosmic variance, and
therefore the value of the lower multipoles cannot be used to exclude
values of $\alpha_-$ greater than but of the order of one.

\section{Comparison with the treatment of perturbations in other models}

\subsection{Perturbations in f(R) theories}
In our previous work \cite{Cortes:2008fy} we found that given an
expansion history parameterized by a modified Friedman equation in a universe
filled by a given component, it was always possible to find a biparametric
family of $f(R)$ theories (see Ref.~\cite{Nojiri:2008nt} for a review)
which had the same homogeneous evolution as
a solution of their equations of motion. We
wonder now if these theories have also the same linearized equations
for the scalar perturbations of the metric.

An $f(R)$ theory is defined by its action and therefore the
equations for the perturbations are uniquely determined. The action is
given by
\be
 S=\frac 1{16 \pi G}\int d^4 x\sqrt{-g}[f(R) + 16\pi G L_m] \, ,
\ee
where $R$ is the curvature scalar and $f(R)$ is an arbitrary function
of this scalar. The equations governing the evolution of the
perturbations will be derived from the Einstein equations of the
$f(R)$ theory,
\be
 f'(R)R^\mu_\nu-\frac12
 f(R)\delta^\mu_\nu+(\delta^\mu_\nu\square-\nabla^\mu\nabla_\nu)f'(R)=8\pi G
 T^\mu_\nu \, .\label{EinfR}
\ee

At the homogeneous level, it is possible to write the Friedman
equations of the $f(R)$ theory,
\begin{widetext}
\begin{eqnarray}
8 \pi G \rho_{(0)} & = & -3(\dH+H^2)f'(\Ro)-\frac 12 f(\Ro) -3 H
f''(\Ro)\dRo \label{fRrho} \, ,\\
-8 \pi G p_{(0)} & = & -(\dH + 3 H^2)f'(\Ro)-\frac 12 f(\Ro)
+f^{(3)}(\Ro)\dRo^2 + f''(\Ro)\ddRo \label{fRp} \, ,
\end{eqnarray}
\end{widetext}

\noindent where
\be
\Ro(t) = -6(\dH +2H^2)
\ee
is the curvature scalar of the Friedman-Robertson-Walker (FRW) metric
(zeroth order in perturbations). The solution of the system of
equations \eqref{fRrho},\eqref{fRp} together with the equation of
state of the dominant component of the stress-energy tensor, defines
$\Ro$ and its derivatives as functions of the Hubble parameter $H$
\cite{Cortes:2008fy}.

If we expand \eqref{EinfR} in powers of the
scalar perturbations of the FRW metric \eqref{newt}, the first order
term gives the linearized equations for the
perturbations. This will be a system of differential equations of
fourth order, but it will be possible to turn it into a system of
differential equations of second order, as we will see below.

If the stress-energy tensor has no shear, the Einstein equation for
$\mu = i\neq \nu = j$ reduces to

\be
f''(\Ro) \delta R +f'(R_0) (\Phi-\Psi) = 0 \label{fRGij} \, ,
\ee
where $\delta R$ is the perturbation of the curvature scalar

\be
\ba{c}
\delta R \,=\, 12(\dH + 2 H^2) \Phi + 6 H \dot{\Phi} + 24 H \dot{\Psi}
\\ + 6 \ddot{\Psi} + \frac 2{a^2}\Delta(\Phi-2\Psi))\, .
\ea \label{delR}
\ee

Eq.~\eqref{fRGij} is a second order differential equation for the scalar
perturbations except in the case  of General Relativity ($f''=0$)
where it becomes an algebraic equation ($\Phi=\Psi$).
Eq.~\eqref{fRGij} can be used to turn the remaining components
of the equations of $f(R)$ theories \eqref{EinfR} into second order
equations.

In the case of adiabatic perturbations in a matter dominated epoch one has

\begin{widetext}
\be
\ba{c}
-8 \pi G \bar{\delta p} =  f'(\Ro)(a^{-2}\Delta(\Phi-2\Psi)+\ddot
\Psi +6 H \dot\Psi + H\dot\Phi +2(\dH + 3 H^2)\Phi -\frac 12 \delta R)
\\ +  f''(\Ro)(-(\dH+3H^2)\delta R -2\dRo \dot \Psi -8 H \dRo
\Psi -4 H \dRo \Phi -2 \ddRo \Phi) - 2
f^{(3)}(\Ro)\dRo^2\Phi  \\
 -\partial^2_t\left[ f'(\Ro)(\Phi-\Psi) \right]-2 H
\partial_t\left[ f'(\Ro)(\Phi-\Psi) \right] = 0 \, .
\ea \label{fRadiab}
\ee
\end{widetext}

If we compare the equations governing the behavior of adiabatic
perturbations in a $f(R)$ theory \eqref{fRGij}, \eqref{fRadiab} with
those coming from the ACM, \eqref{Phi},\eqref{Psi}, we find that they
describe very different
behaviors. In one case we have a system
of two coupled second order differential equations for $\Phi$ and
$\Psi$. In the ACM case we got a single second order differential
equation and an algebraic equation between the two gravitational
potentials. For a given $f(R)$ theory with an associated $g(H)$
homogeneous behavior, we find that the behavior of linearized
perturbations differs from the one defined by the linearized
perturbations in the ACM.  This means that $f(R)$ theories break in
general our assumption in the third section (there are terms with more than
two derivatives in the equations for the perturbations, and therefore
the behavior of perturbations in General Relativity
is not recovered when $H_+ \gg H \gg H_-$ for all modes).

Let us see this in a simple example. We will choose the easiest
biparametric family of $f(R)$ theories: the one that, under matter
domination, gives the same background evolution as General Relativity
without a cosmological constant \cite{Cortes:2008fy},

\be
f(R)=R+c_1 \vert R \vert^{\frac 1{12}(7-\sqrt{73})}+c_2 \vert R
\vert^{\frac 1{12}(7+\sqrt{73})} \, \label{fRGR}.
\ee
These kind of models were introduced in Ref.~\cite{Nojiri:2003ft}.

Let us first check if the solutions of the equations for the
perturbations in General Relativity in the presence of pressureless
matter,

\be
\begin{array}{ccl}
\Psi = \Phi & = & const \, ,\\
\Psi = \Phi & \propto & t^{-5/3} \, , \label{solsGR}
\end{array}
\ee
are solutions of the equations of motion in the case of the
biparametric family of $f(R)$ theories (\ref{fRGR}).
The easiest is to verify if Eq.~\eqref{EinfR} with $\mu = i \neq \nu = j$,
\be
\begin{array}{c}
2 f''(\Ro) (3 H^2 \Phi + 3 H \dot{\Phi} + 12 H \dot{\Psi} + 3
 \ddot{\Psi} + \frac 1{a^2}\Delta(\Phi-2\Psi))\\
 +f'(\Ro) (\Phi-\Psi) = 0 \label{fRGijGR} \, ,
\end{array}
\ee
where $\Ro (H) = -3 H^2$ is the value of the homogeneous curvature
scalar as a function of $H$ in this family of theories, is also
fulfilled by the solutions (\ref{solsGR}). The result is obviously
not.

The second order differential equations for the perturbations in the
$f(R)$ theories (\ref{fRGR}) in the presence of matter can be found by
substituting (\ref{fRGijGR}) and its derivatives into the other
equations of the system \eqref{EinfR}. The equation for the pressure
perturbations then gives
\be
\begin{array}{c}
-f'(\Ro)[\ddot{\Psi}+\ddot{\Phi}+ 4 H
  \dot{\Psi}+4H\dot{\Phi}+3H^2\Phi] \\
-27H^3f''(\Ro)\dot{\Phi}+\frac{f(\Ro)}2(\Psi-3\Phi)\,=\,8\pi G
  \bar{\delta p} \, = \, 0 \, \label{fRGiiGR} \, .
\end{array}
\ee
It is obvious that the $i\neq j$ equation of the cosmologic
perturbations in General Relativity,
\be
\Phi - \Psi = 0
\ee
is not recovered from (\ref{fRGij}) in the $H \gg H_-$ limit for all
$k\ll H_+$. In fact the deviation would be significant for modes with
$k \gtrsim H a \left(\frac{H}{H_-}\right)^{\frac 1{12}(7-\sqrt{73})}
\sim H a$. Therefore, the assumption we have made in order to derive
the equations for the perturbations in the ACM is broken.

It is a known problem that $f(R)$ theories are unable to pass
cosmological and astrophysical tests involving perturbations \cite{Chiba:2003ir},
unless the function involved is properly fine-tuned. In particular, if the
conformal equivalence between $f(R)$ theories and scalar-tensor theories is
used, it is necessary that the effective mass acquired by the new scalar
degree of freedom is unnaturally large \cite{Faulkner:2006ub}.
Some $f(R)$ theories that pass cosmological and solar system tests have been
proposed \cite{Hu:2007nk,Nojiri:2007as,Cognola:2007zu}.
However, it is also subject of debate if the conformal equivalence can
be used in order to extract physical predictions from this models, especially
predictions which involve perturbations \cite{Carloni:2009gp}. A mathematically
rigorous treatment of perturbations in $f(R)$-gravity can be found in
Ref.~\cite{Carloni:2007yv}.

\subsection{Quintessence Models}

Another class of models used to describe the acceleration of the
universe are those in which a so called quintessence field, typically
of scalar type $\varphi$, is added as a component of the universe
\cite{Wetterich:1987fm}. The only functional degree of freedom in most
models is just the scalar potential $V(\varphi)$, which can be tuned
to fit the homogeneous expansion of the universe. In our previous work
\cite{Cortes:2008fy} we found the correspondence between a given
homogeneous evolution parameterized by $g(H)$ and the potential for
the quintessence which drives under General Relativity this
evolution. We now wonder if the behavior of perturbations in
quintessence models also resembles the behavior in ACM.

The field action is given by

\be
S_\varphi = \int d^4 x \sqrt{-g} \left[ g^{\mu\nu}\partial_\mu \varphi
  \partial_\nu \varphi -V(\varphi) \right] \, . \label{quint}
\ee

Let us assume that the field can be split in two components: one which
is only time dependent and which drives the homogeneous evolution of
the universe, $\varphi_{(0)} (t)$, and a small perturbation which is
inhomogeneous, $\delta\varphi({\bm x},t)$. We will consider linearized
perturbations of the metric, the field, and the other components of
the universe. The resulting linearized stress-energy tensor of the
field is

\begin{eqnarray}
\delta T^\varphi_{00} & = & \dot{\varphi}_{(0)}
\dot{\delta\varphi}-\Phi V(\varphi_{(0)}) + \frac 12
V'(\varphi_{(0)})\delta\varphi \, , \\
\delta T^\varphi_{0i} & = & \dot{\varphi}_{(0)} \partial_i \delta\varphi \, ,\\
\delta T^\varphi_{ij} & = & a^{-2}
\delta_{ij}\left[\dot{\varphi}_{(0)}^2 (\Phi-\Psi)
  +\dot{\varphi}_{(0)} \dot{\delta\varphi} \right. \nonumber \\
& & \left. +\Psi V(\varphi_{(0)}) - \frac 12 V'(\varphi_{(0)})\delta\varphi
\right] \, .
\end{eqnarray}

By virtue of the Einstein's equations, it is always possible to turn a
modification in the Einstein tensor into a new component of the
stress-energy momentum tensor of the sources of the gravitational
field. We wonder if it is possible to account for the modification of
the effective Einstein tensor described in the third
section with a new component described by this quintessence
field. However, it is not hard to see that the effective Einstein
tensor that we are proposing has a modified $i\neq j$ component, while
the $i\neq j$ component of the stress-energy tensor of the quintessence field
$\varphi$ is zero. Therefore, it is not possible to describe the
evolution of perturbations in the ACM as driven by an effective scalar
field component~(\ref{quint}).

\subsection{Dark Fluid Models}

A deformation of the gravitational physics can be also made equivalent
to the addition of a non-standard fluid component to the cosmic pie at
the homogeneous level. The fluid component used to explain the present
accelerated expansion of the universe is typically taken to be a
perfect fluid with large negative pressure.

One of the most popular parameterizations of this fluid is the so
called equation of state $\omega = p / \rho$ \cite{Turner:1998ex}. The
simplest models are $\omega = -1$, which is equivalent to a
cosmological constant, or a constant $\omega$, but recently a possible
time dependence of $\omega$ has been considered
\cite{Bassett:2004wz}. Needless to say, these {\it ad hoc}
parameterizations are subsets of the more general equation of state of
a perfect fluid, $p = p(\rho,s)$, where $s$ is the entropy density. It
is also questionable why must we restrict ourselves to perfect fluids
and not include a possible anisotropic stress tensor, possibly
depending on the energy and entropy densities
\cite{Capozziello:2005pa}.

As in the previous subsection, it is always possible to use the
Einstein's equations to turn a modification in the gravitational
physics into a new fluid component of the universe. In the case of the
equations derived in the third section, the
equivalent Dark Energy Fluid would have the following properties:

{\small
\begin{eqnarray}
 8\pi G \bar{\delta \rho}_X & = & (2H-g')\left[\frac 1{a^2 H}\Delta
 \Psi -3(H\Phi + \dot{\Psi})\right], \\
 8\pi G \bar{\theta}_X & = & \frac{(2H-g')}H(H\Phi + \dot{\Psi}), \\
 8\pi G \Pi_X & = & \frac
 1{a^2}\left\{\frac{2H-g'}{2H}(\Phi-\Psi)-\frac{\dot{H}}{2H^3}(g'-Hg'')\Psi\right\},
 \\
8\pi G \bar{\delta p}_{X} & = & (2H-g')(3H\Phi +3\dot\Psi-\frac{\dot
 H}{H^2}\dot \Psi + \frac{\dot H}{H}\Phi + \dot \Phi + \frac 1H
 \ddot\Psi)\nonumber \\
& & +\frac{\dot H}{H}(2-g'')(H\Phi+\dot\Psi)-8\pi G \Delta\Pi_X.
\end{eqnarray}
}

The resulting fluid is not an ideal fluid ($\Pi_X = 0$) or even a
Newtonian fluid ($\Pi_X \propto \bar\theta_X$). With the use of
\eqref{gh}, \eqref{Phi} and \eqref{Psi}, it will be possible to write
$\bar{\delta\rho}_X$ and $\bar{\delta p}_X$ as a combination of $\bar
\theta_X$, $\Pi_X$ and their derivatives,

\begin{eqnarray}
 \bar{\delta\rho}_X & = & f_1(\rho_{(0)X}) \Delta \Pi_X +
 f_2(\rho_{(0)X}) \bar\theta_X \, ,\\
\bar{\delta p}_X & = & f_3(\rho_{(0)X})\Pi_X -\Delta\Pi_X +
 f_4(\rho_{(0)X})\bar\theta_X \, .
\end{eqnarray}

These very non-standard properties show that, although it is formally
possible to find a fluid whose consequences mimic the ones of such a
modification of the gravitational physics, this fluid is very exotic.

\section{Summary and Conclusions}

In view of recent data, an updated comparison of cosmological
observations with a phenomenological model proposed in a recent work
has been presented.

An extension of this phenomenological model (ACM)
beyond the homogeneous approximation has been introduced allowing us to
describe the evolution of scalar perturbations at the linear level.

A comparison with the
spectrum of thermal fluctuations in CMB has been used to explore the
possibility to determine the parameters of the ACM through its role in
the evolution of scalar perturbations. The results of this comparison
does not further restrict the parameters of the model, due to the masking
of the associated late ISW effect by the cosmic variance. However, better
measurements of the position of the third acoustic peak should improve
the constraints significatively.

It has been shown that the equivalence of different formulations of
the accelerated expansion of the universe in the homogeneous
approximation is lost when one considers inhomogeneities. In
particular we have shown that the general structure of the evolution
equations for scalar perturbations in the ACM differs from the
structure of the equations corresponding
to modified $f(R)$ theories of gravity, to quintessence models or to a
dark fluid with standard properties.

The possibility of going beyond
the linearized approximation for the scalar perturbations and to
consider vector and tensor perturbations will be the subject of a
future work.

\acknowledgments

We would like to thank Marco Bruni for inspiration. We would also like
to thank Valerio Faraoni, Daniel G. Figueroa, Troels Hagub\o{}lle,
Sergei Odintsov and
Javier Rubio for fruitful discussions. J.I. also thanks the
hospitality of the Benasque Center for Science and the Scuola
Internazionale Superiore di Studi Avanzati (SISSA) during the
development of this work.

This work has been partially supported by CICYT (grant
FPA2006-02315) and DGIID-DGA (grant2008-E24/2). J.I. acknowledges a FPU
grant from MEC.

\appendix

\section{Appendix: Covariant Perturbation Equations for the ACM}
The code of CAMB \cite{CAMB} makes use of the equations for the
perturbations of the metric in the covariant approach. The quantities
can be computed in a given ``frame'', labeled by a 4-velocity
$u^\mu$. In particular, CAMB uses the dark matter frame, in which the
velocity of the dark matter component is zero (the dark matter
frame).
Furthermore, it parameterizes the time evolution with the conformal
time $a(\tau)d\tau = dt$.

In order to apply CAMB to the ACM model it is necessary to
identify frame invariant quantities, and then relate them to
gauge invariant quantities \cite{Gordon:2002gv}. The following
comoving frame quantities are used in the CAMB code: $\eta$ (the
curvature perturbation), $\sigma$ (the shear scalar), $z$ (the
expansion rate perturbation), $A$ (the acceleration, $A=0$ in the dark
matter frame), $\phi$ (the Weyl tensor perturbation), $\chi^{(i)}$
(the energy density perturbation of the species i), $q^{(i)}$ (the
heat flux of the species i), and $\Pi^{(i)}$ (the anisotropic
stress of the species i). All of these quantities are defined in
\cite{Challinor:1998xk}.  These variables are related to the gauge
invariant variables via the following dictionary,

\bea
-\frac \eta 2 -\frac{\mH\sigma}{k} & \equiv & \Psi \, ,\\
- A + \frac{\sigma'+\mH\sigma}{k} & \equiv & \Phi \, ,\\
\Ji + \frac{\rho'\sigma}{k} & \equiv & \bar{\delta\rho} \, ,\\
\rho q + (\rho + p)\sigma & \equiv & \frac k a (\rho + p) \bar{\theta}\, ,
\eea
where prime denotes derivatives with respect to conformal time,
except when acting on $g(H)$ where it denotes a derivative with
respect to the Hubble rate $H$, and $\mH = a'/a = aH$. On the other hand
$z = \sigma + \frac{3}{2k}(\eta'+2\mH A)$ and $\phi = (\Phi +
\Psi)/2$. The anisotropic stress $\Pi$ is already frame invariant.

Written in the dark matter frame, the equations for the scalar perturbations read
\begin{widetext}
\be
\begin{array}{rcl}
g'\left(\frac{k^2 \eta}{2 \mH} + k z\right) & = & 8\pi G a \Sigma_i \Ji^{(i)} \, ,\\
\frac{g' k^2}{3\mH}(\sigma - z) = -\frac{g'k}{2\mH} \eta' & = & 8\pi G
a \Sigma_i \rho_i q^{(i)}\, ,\\
-\mH k g'(z' + \mH z) & = & 8 \pi G a \Sigma_i\left[
  \mH^2 (1+3c_s^{(i)2}) - (\mH'-\mH)^2(1-\frac{\mH g''}{a g'}) \right]\Ji^{(i)}\, ,
\\
\frac {g'}\mH \left(\frac{\sigma'+\mH \sigma}{k}-\phi
\right)-\frac{\mH'-\mH^2}{2 \mH^3}(g'-\mH g''/a) \left(\frac\eta
2+\frac{\mH\sigma}k \right)& = & - 8\pi G a \Sigma_i\Pi^{(i)}/k^2 \, .
\end{array}
\ee
The following combination of the constraint equations is also useful:
\be
k^2\phi = -\frac{8\pi G a \mH}{g'}\Sigma_i\left[ \Pi^{(i)}
  +(1-\frac{\mH'-\mH^2}{2 \mH^2}(1-\frac{\mH g''}{a g'}))(\Ji^{(i)}
  +3\mH \rho_i q^{(i)}/k)\right]\, .
\ee
\end{widetext}

These equations are plugged into the Maple files provided with the
CAMB code and run to get the ISW effect.

\end{document}